# A new lagrangian particle method to describe turbulent flows of fully compressible ideal gases

B. Ivancic
CFD2k-Software Development and Engineering Corporation (CFD2k-SDEC)
5210 Windisch ; Switzerland
Email: info@cfd2k.eu
(Dated: February 25, 2008)

## Abstract

There are several approaches to describe flows with particles e.g. Lattice-Gas Automata (LGA), Lattice-Boltzmann method (LBM) or smoothed particle hydrodynamics (SPH). These approaches do not use fixed grids on which the Navier-Stokes equations are solved via e.g. finite volume method. The flow is simulated using a multitude of particles or particle density distributions, which interacts and due to statistical laws and an even more fundamental approach than the Navier-Stokes equation, the averaged flow variables can be derived. After a short summary of the most popular particle methods the new DMPC (Dissipative Multiple Particles Collision) approach will be presented. The DMPC-model eliminates some of the weak points of the established particle methods and shows high potential for more accurate CFD solution especially in areas where standard CFD tools still have problems (e.g. aero-acoustics). The DMPC-model deals with discrete circular particles and calculates the detailed collision process (micro scale) of several overlapping particles. With thermodynamic, statistical and similarity laws global (large scale) flow variables can be derived. The model is so far 2d and the particles can move in every direction in the 2d plane depending on the forces acting on it. The possible overlap between neighbouring particles and multi-particle interactions are important features of this model. A freeware software is developed and published under www.cfd2k.eu. There the executable, the user guide and several exemplary cases can be downloaded.

## Introduction

Standard CFD solvers work with fixed grids (Euler approach). On this grid the governing equation are solved numerically with a variety of approximate schemes, which are available nowadays [1]. The most general governing equations are the Navier-Stokes equations but models like Euler-equations for inviscid flows are preferably used if this simplification is possible. In fluid dynamics for Euler approaches most popular solver methods are the finite volume methods, finite difference methods and finite element methods.

Particle methods work principally different. They use no fixed grid (mesh free method). The particles are distributed over the whole domain and interaction between the particles combined with statistical approaches enables a derivation of averaged flow field variables. Particle methods do not solve the Navier-Stokes (N-S) equations directly because the main difference is that N-S approaches start with a mathematical description of the flow at a continuum level. Particle methods work on a more fundamental i.e. kinetic level. The most popular particle methods are the Lattice-Boltzmann and the SPH (smoothed particle hydrodynamic) methods. These two methods will be now shortly summarised together with their advantages and disadvantages before the new DMPC method with its benefits is presented.

The Lattice-Boltzmann method [2, 3] was developed at the end of the 80's. It is based on the simulation of strongly simplified particle micro dynamics. The simulation on particle level needs a very small amount of computational resources per particle due to the simple inner structure. For this reason this method is appropriate for the simulation of very complex geometries like porous media etc. The Lattice-Boltzmann method has its origin in statistical physics and the governing equations are the Boltzmann equations:

$$\frac{\partial f}{\partial t} + v \frac{\partial f}{\partial x} + F \frac{\partial f}{\partial v} = \Omega \tag{1}$$

The Boltzmann equation is an evolution equation for a single particle probability distribution function $f(x,v,t)$ where $x$ is the position vector, $v$ the particle velocity vector, $F$ is an external force and $\Omega$ is a collision integral. The lattice Boltzmann method discretizes equation (1) by limiting space to a lattice and the velocity space to a set of discrete particle velocities $v_i$. The discretised Boltzmann equation, which is the Lattice Boltzmann equation, can be transformed in the N-S equations. The Lattice Boltzmann method considers particle distributions that are located on lattice nodes and not individual particles. The general form of the lattice Boltzmann equation is:

$$f_i(x + v_i \cdot \Delta t, t + \Delta t) = f_i(x,t) + \Omega_i \tag{2}$$

Where $f_i$ can be interpreted as the concentration of particles that flows with the velocity $v_i$. With this discrete velocity the particle distribution travels to the next lattice node, which is reached in one time step $\Delta t$. There are several approaches existing to model the collision integral $\Omega_i$. Very popular is the Bhatnagher-Gross-Krook method [4]. Here the particle distribution after propagation is relaxed towards the equilibrium distribution $f_i^{eq}(x,t)$ as:

$$\Omega_i = \frac{1}{\tau}\left(f_i(x,t) - f_i^{eq}(x,t)\right) \tag{3}$$

Here $\tau$ is the so-called relaxation parameter, which determines the kinematic viscosity $\nu$ of the simulated fluid, according to the following equation:

$$\nu = \frac{2 \cdot \tau - 1}{6} \tag{4}$$

Local flow variables can be derived by the first two moments of the particle distribution function $f_i$. The first moment of $f_i$ is the local density $\rho$, as it is defined in the following equation:

$$\rho(x,t) = \sum_i f_i(x,t) \tag{5}$$

The second moment is the local flow velocity vector:

$$u(x,t) = \frac{\sum_i f_i(x,t) \cdot v_i}{\rho(x,t)} \tag{6}$$

The equilibrium distribution $f_i^{eq}(x,t)$ is a function of the local density $\rho(x,t)$ and the local flow velocity vector $u(x,t)$. This equilibrium density function can be calculated as follows:

$$f_i^{eq}(\rho,u) = t_p \cdot \rho \cdot \left(1 + \frac{v_i.u}{c_s^2} + \frac{(v_i.u)^2}{2 \cdot c_s^4} - \frac{u.u}{2 \cdot c_s^2}\right) \tag{7}$$

In equation (7) $c_s$ is the speed of sound. Depending on the dimensionality of the space, which has to be simulated there are various sets of velocity and corresponding equilibrium distributions that can be used. For a 2d simulation and 9 velocity sets (Lattice: D2Q9) the corresponding weightings are $t_0=4/9$, $t_1=1/9$ and $t_2=1/36$.

In the lattice-Boltzmann approach the particles are only located on discrete lattice nodes. The particles are moving from one lattice node to the next depending on its discrete velocity. This phase is called the propagation phase, which has to be distinguished from the collision phase. This happens when the particle is placed on a certain lattice node. In the collision phase the particles gets new velocities due the equations described above. The particles are inherently trying in each collision phase to come closer to the equilibrium distribution.

Before the advantages and disadvantages of this approach will be discussed in order to compare it with other particle methods and the new DMPC approach, it will be now the SPH method discussed which is a further alternative in particle methods.

The SPH is a numerical method to solve the hydrodynamic equations [5, 6]. This method is also used e.g. in Astrophysics [7, 8] and tsunami prediction [9]. The SPH approach trenches the fluid in discrete elements (particles), which are randomly distributed over the domain. The mean distance between these particles is the smoothing length $h$, which is the most important parameter of this method. Between the particles the fluid properties are smoothed by the so-called kernel-function. Every flow variable (e.g. local density $\rho$) is calculated by the sum over all particles, which are located within two smoothing lengths. Each particle possesses a fraction (scalar) of the local flow variable. Due to this process the original partial differential equations, which describes the hydrodynamics, are transformed to the much simpler ordinary differential equations. It has to be emphasized here that SPH is a very empirical approach. This means that SPH approaches are often used because they are working well but there is no fundamental mathematical derivation.

The formal deduction of the SPH methods is done either by a Lagrange Function or via an integral-interpolation. At the integral-interpolation method one assumes the identity for a variable $A_1$ as follows:

$$A_1(\vec{r}) = \int A(\vec{r}') \cdot \delta(\vec{r}-\vec{r}') \cdot d\vec{r}' \approx \int A(\vec{r}') \cdot W(\vec{r}-\vec{r}',h) \cdot d\vec{r}' \qquad (8)$$

In the upper equation $\delta(\vec{r}-\vec{r}')$ is the Dirac delta-distribution. This delta-distribution will be approximated with a kernel function $W(\vec{r}-\vec{r}',h)$, where $h$ is the smoothing length. In order to achieve the validity of this approximation also in the extreme case when the smoothing length goes to infinity, one has to demand a normalisation and the identity of the δ-function with the kernel-function for h→0. This means the following equations must be fulfilled:

$$\begin{aligned} \int W(\vec{r}-\vec{r}',h) \cdot d\vec{r}' &= 1 \\ \lim_{h \to 0} W(\vec{r}-\vec{r}',h) &= \delta(\vec{r}-\vec{r}') \end{aligned} \qquad (9)$$

To get a discretisation in mass elements it is useful to divide with the density. In the extreme case of infinite number of particles, which are infinite small the sum goes over in the following integral:

$$A_S(\vec{r}) = \lim_{h \to 0} \int \frac{A(\vec{r}')}{\rho(\vec{r}')} \cdot W(\vec{r}-\vec{r}',h) \cdot \rho(\vec{r}') \cdot d\vec{r}' \propto \sum_{b=1}^{N} m_b \cdot \frac{A_b}{\rho_b} \cdot W(\vec{r}-\vec{r}',h) = A_b \qquad (10)$$

Of course the computers can treat only a limited number of particles, what means that for the numerical approach the right side of the upper equation is relevant. The right side of equation (10) is the basic equation of SPH. The variable A is calculated by a sum over all particles. The variable $A_s$ which is depending on $r$ is transformed to a scalar $A_b$ which is multiplied with the kernel. This leads to a strong simplification of the differential equations because the derivation is not affecting any more the variable but only the kernel, which is shown in the following equation:

$$\nabla A(\vec{r}) = \sum_b m_b \cdot \frac{A_b}{\rho_b} \cdot \nabla W(\vec{r}-\vec{r}',h) \qquad (11)$$

The kernel is the most important function within the SPH method. One can compare the different kernels with the various finite difference schemes for the grid approaches. For a better understanding of the SPH approach it is helpful to define a kernel similar to the Gaussian curve:

$$W(\vec{r}-\vec{r}',h) \propto e^{-\left(\frac{(\vec{r}-\vec{r}')^2}{h^2}\right)} \qquad (12)$$

Numerically this approach has some difficulties, because the coverage of the kernel is clearly fixed and increasing the distance $r$ over a certain value leads to a kernel, which is zero. This results in a strong limitation of the neighbouring particles, which are taken into account. There are several kernels available, which are used for different tasks. There are no clear rules for the choice of the kernel; mostly the selection is done by trial and error. Experience plays here an important role. As mentioned before SPH is a strongly empirical approach.

Apart from the kernel the smoothing length *h* is a very important parameter. The smoothing length determines the resolution of the simulation and due to this it has a strong impact on the accuracy and computational costs. Together with the kernel the smoothing length determines the number of neighbouring particles, which are taking into account. Typically there are several dozens particles taken into account for the calculation of one variable. To achieve more accurate results the smoothing length is scaled with the fluid density:

$$h \propto \frac{1}{\langle \rho \rangle^{\frac{1}{k}}} \qquad (13)$$

Where *k* defines the dimensionality. The fluid density can be calculated as follows:

$$\langle \rho \rangle = \frac{1}{n} \cdot \sum_b \rho_b \qquad (14)$$

The number of particles is given by *n* and $\rho_b$ is the density of one discrete particle. In modern codes the smoothing length is calculated time dependent, as follows:

$$\frac{dh_a}{dt} = -\left(\frac{h_a}{k \cdot \rho_a}\right) \cdot \frac{d\rho_a}{dt} \qquad (15)$$

This equation ensures a fine resolution in areas where the density is high, while in regions of small density the smoothing length gets smaller. By this approach the computational costs decreases with same accuracy level.

The most popular particle (mesh free) methods in fluid dynamics are the two methods, the Lattice-Boltzmann and the SPH, which are presented here. For completeness it has to be mentioned that there are apart from these other particle methods e.g. discrete vortex methods [10] and various kinds of approaches, which can be summarised under the generic term particle-in-cell methods (PIC) [11]. PIC methods are often used in plasma physics [12]. To keep the scope of the paper in an acceptable range these approaches will not be discussed in detail here, but general overviews of these methods are given in [13, 14, 15, 16, 17].

# Comparison of the assets and drawbacks between the usual particle approaches

One of the main disadvantages of all particle methods is the inherent statistical noise of these methods. Increasing the number of particles can reduce the noise but this leads to higher computational costs. The need of a high number of particles to reduce the noise and to achieve better accuracy disprove the common opinion that particle methods need less computational efforts than traditional CFD methods like finite volume or finite element methods. The dependency of the calculating effort on particle number N can be given for optimised algorithms as N*logN. In this context it has to be mentioned as advantage of particle methods that its parallelisation is in general simpler as for standard grid-methods. Another general advantage of particle methods is their automatically fulfilled mass conservation. This is an inherent feature of lagrangian methods. Additionally to that particle methods are comparatively easy to develop and to implement in a code. Especially for the SPH method the exchange of different kernels is quite simple. Particle methods are generally much more robust than grid methods i.e. one can achieve almost for all cases a converged solution. This can also be a drawback because especially for the SPH method, one can get results also by using completely wrong kernels for the regarded case. These results are of course also completely wrong and this has to be recognised. The error analysis for particle methods is often very difficult and only possible by comparative studies with other approaches. Additional disadvantages of the SPH method is, that it is highly dispersive and the treatment of discontinuities is very difficult, because structures and scales which are smaller than the smoothing length will be smoothed. The Lattice-Boltzmann method, which is very popular in the simulation of complex fluid systems, has difficulties in high Mach number flows in aerodynamics. Additionally to that heat transfer processes causes problems for this method.

The new DMPC method possesses more or less all of the mentioned assets. Furthermore some of the drawbacks of the discussed particle methods are solved or at least mitigated. Especially the simulation of discontinuities (e.g. rarefaction waves or shockwaves) has been demonstrated. In addition to that the new DMPC method shoes excellent results even for a relative small number of particles.

## Main features and mathematical description of the DMPC-model

In the DMPC method the whole domain is filled with circular particles, which can overlap each other. The current version of the test-code work with constant particle diameter, but it is possible to extend the model also to variable particle diameter or the even more sophisticated approach using so-called moving Voronoi-particles. Of course the equations are getting then more complicated, but this can improve some current difficulties like the exact modelling of the transport processes (e.g. turbulent diffusion). The particles represent artificial flow elements, which are moving in space and time. They are carrier of certain variables and possess certain thermodynamic characteristics (dissipative and dispersive). The particles are interacting with neighbouring particles and boundaries. With the help of thermodynamic, statistical and similarity laws it is possible to derive averaged flow field variables for a local conglomeration of particles. The Navier-Stokes equations are not solved directly but can be derived with the particle collision and interaction laws. The particles can also be regarded as a kind of self-adaptive sub-grid cells (micro-scale), which are moving due to multiple particle interaction. Several dozens of these sub-grid cells are inside one single so-called rigid "post-processing-grid (ppr-grid)" (macro-scale). The averaged macroscopic flow field variables are interpolated on this fixed ppr-grid, which is directly visible for the user. On this macroscopic ppr-grid the post-processing of the data is done. At the current stage of the test code CFD2k, only hexahedral ppr-grids are allowed, but it is not a problem to extend the code to more flexible tetrahedral ppr-grids.

All particles possess 6 variables, which fully describe the thermodynamic state of the particle. These variables are: The position vector $\vec{r}$ (x, y), velocity vector $\vec{V}$ (Vx, Vy), mass $m$ and total temperature $T_{tot}$. The movement of each particle is described by the following Newton's equations of motion:

$$\frac{\partial \vec{r}_i}{dt} = \vec{V}_i$$
$$\frac{\partial \vec{V}_i}{dt} = \sum_i \nabla e_{pot,i,Nb}$$
(16)

In equation (16) $e_{pot,i,Nb}$ characterizes the specific potential energy between the particle $i$ and a neighbouring particle $Nb$. The potential energy, its meaning and definition, will be explained later in more detail. This approach assumes that every individual particle is in thermodynamic equilibrium. With the usage of the ideal gas law, the particle pressure can be derived:

$$P = \rho \cdot R \cdot T \tag{17}$$

The static temperature of the particle is determined with gas dynamic laws:

$$T = T_{tot} - \frac{\vec{V}^2}{2 \cdot c_p} \tag{18}$$

The current approach works with constant heat capacity $c_p$. But it should not be an issue to extend it to more realistic, temperature dependent heat capacities. The core of the DMPC model is the description of the particle interaction or collision process. This interaction process is applied simultaneously to all neighbouring (overlapping) particles around the regarded central particle and comprehends a dissipative and a repulsion or dispersive part. At the same time relaxation processes between the interacting particles are modelled which are mass and energy diffusion.

$$\frac{\partial c}{\partial t} = \nabla \cdot (D \nabla c) \tag{19}$$

In equation (19) D is the mass diffusion coefficient for the mass relaxation process or the conductibility of temperature for the energy relaxation process.
The dissipative part of the multi-particle collision process can also be interpreted as a kind of momentum relaxation of the participating particles. In the further model description the particle momentum vector will be used which is defined for one particle with the index $i$:

$$\vec{I}_i = m_i \cdot \vec{V}_i \tag{20}$$

For the dissipative part, first the mean momentum vector of all $N$ interacting particles has to be calculated.

$$\bar{\bar{I}} = \frac{\sum_i m_i \cdot \vec{V}_i}{N} \qquad (21)$$

The dissipation process can here be interpreted with the inherent affinity of every particle to harmonize its velocity towards this mean velocity. If at the end of the collision process all particles would have the same mean velocity than the dissipation would be complete, but this in reality can not happen because of two reasons: first there is a dissipation scaling parameter $a$ (see eq. 29) which is always <= 1 and secondly the dispersive part of the collision process which affects all particle velocities in that way that they tend to diverge. The differential momentum vector of a particle to the mean momentum vector can be written as:

$$\vec{I}_i + d\vec{I}_{i,dissip} = \bar{\bar{I}} \qquad (22)$$

It can be shown that momentum conservation:

$$\sum_i d\vec{I}_{i,dissip} = 0 \qquad (23)$$

is always fulfilled for all interacting particles. An important feature is that these differential momentum vectors can now be scaled with a constant dissipation scaling parameter $a$ and the momentum conservation will always be fulfilled. The dissipation scaling parameter is a model variable and determines the strength of the dissipation process. During the dissipation kinetic energy is transferred between the particles, which is proportional to the potential energy of the particle arrangement. The potential energy of the particle arrangement is increasing when the particles are closer together i.e. when the overlapping volume is increasing. Simultaneously to the calculation of the dissipative part the dispersive part is determined, too. Here every neighbouring particle gets differential momentum vector in the direction away from the central particle:

$$d\vec{I}_{i,Nb,dispersive} = m_{i,Nb} \cdot \frac{\vec{r}_{i,Nb} - \vec{r}_{Cent}}{|\vec{r}_{i,Nb} - \vec{r}_{Cent}|} \qquad (24)$$

In order to keep momentum conservation also for this part of the collision model the central particle needs to get a differential momentum vector which compensates the sum of all neighboring differential momentum vectors:

$$d\vec{I}_{cent,dispersive} = -\sum_i d\vec{I}_{i,Nb,dispersive} \qquad (25)$$

Both processes, the dissipative and the dispersive part, transfer kinetic energy between the particles and if boundaries are involved also between the boundaries. The distribution of these transferred kinetic energies determines the final state after the collision process. There is no fundamental mathematical derivation how these energies have to be distributed. This model was developed more intuitive and with trial and error i.e. the energy transfer model, which shows the best results is finally implemented. But there are some rules, which must be fulfilled. First the transferred energy in one collision process cannot be higher then the potential energy of the overlapped particle arrangement i.e. the regarded local particle conglomeration. Or secondly: as closer the particles are together i.e. the potential energy of the overlapped particle arrangement increases as stronger gets the dispersive part of the collision process compared to the dissipative part. The potential energy $E_{pot}$ of a central particle, which is overlapped by $N$ neighboring particles, can be calculated as follows:

$$E_{pot} = \sum_i \frac{P_{cent} + P_{i,Nb}}{2} \cdot V_{ov,i,Nb} \tag{26}$$

The overlapping volume between 2 neighboring particles can be calculated as follows:

$$V_{ov,i,Nb} = \frac{D^2}{2} \cdot \left( a\cos\left(\frac{D}{d}\right) - \sqrt{1 - \left(\frac{D}{d}\right)^2} \right) \tag{27}$$

Where $D$ is the particle diameter and $d$ is the distance between 2 neighboring particles.

The transferred kinetic energy of one particle depends on the differential momentum vector but in 3 different ways. If the differential momentum causes a particle acceleration the transferred energy is:

$$dE_{kin,i} = m_i \cdot \left( \vec{V_i} \cdot \frac{d\vec{I_i}}{m_i} + 0.5 \cdot \left(\frac{d\vec{I_i}}{m_i}\right)^2 \right) \tag{28}$$

If the particle is decelerated i.e. $V_{x,i} \cdot dI_{x,i} < 0$ or $V_{y,i} \cdot dI_{y,i} < 0$, the equation looks like this:

$$dE_{kin,i} = m_i \cdot \left| \vec{V_i} \cdot \frac{d\vec{I_i}}{m_i} + 0.5 \cdot \left(\frac{d\vec{I_i}}{m_i}\right)^2 \right| \tag{29}$$

Upper equation is only valid if the particle velocity decreases but it does not change the direction signum. If additionally the direction signum is changed the transferred energy of a particle must be calculated in this way:

$$dE_{kin,i} = m_i \cdot \left( \frac{d\vec{I_i}}{m_i} \cdot \left( \vec{V_i} + 0.5 \cdot \frac{d\vec{I_i}}{m_i} \right) + \vec{V_i} \cdot \vec{V_i} \right) \tag{30}$$

The two parts of the collision process i.e. the dissipative and the dispersive are calculated simultaneously. Superposition principles can be used to overlay both vectors to calculate the final differential momentum vector representing the final collision process:

$$d\vec{I}_{collision} = a \cdot d\vec{I}_{dissip} + b \cdot d\vec{I}_{dispersive} \tag{31}$$

The model variable *a* is the dissipation scaling parameter and *b* the dispersion scaling parameter. These model parameters are calculated in that way that several constraints regarding the transferred energies are fulfilled. The partitioning between the transferred energies for the dissipative and the dispersive part depend on the arrangement of the interacting particles and their mean distance. If this distance gets smaller the dispersive part gets stronger and also the total transferred energy is higher because the potential energy of the particle agglomeration gets higher. There is a sophisticated algorithm for this energy distribution model, which was developed with trial and error because no fundamental mathematical derivation could be found. But the final algorithm produces results, which fits best with validation data. This approach is fully mass, momentum and energy conserving and it obeys the Navier-Stokes equations. But there are still some difficulties for special particle arrangements e.g. very rare particle distribution in wakes. In such cases additional model features has been developed which works with particle insertion or particle merging in areas where it is necessary but keeping all conservation laws fulfilled. The latest model improvement is the replacement of the overlapping circular particles by so-called moving Voronoi-particles. This model extension mitigates some problems but there is still a lot of development work necessary to optimize it further.

## Results and Validation

The developed tool was tested and validated on simple academic cases. Several exemplary results are shown in www.cfd2k.eu. There are also animations displayed of transient flow behaviour, start-ups and pressure waves travelling through the domain. Here results and comparisons are shown for backward facing step simulations. This validation case was taken from the ERCOFTAC classic database (case #31, see http://cfd.mace.manchester.ac.uk/ercoftac/) and DNS simulations where done by Moin [18]. Figure 1 shows a contour plot of normalized axial velocity and the ppr-grid.

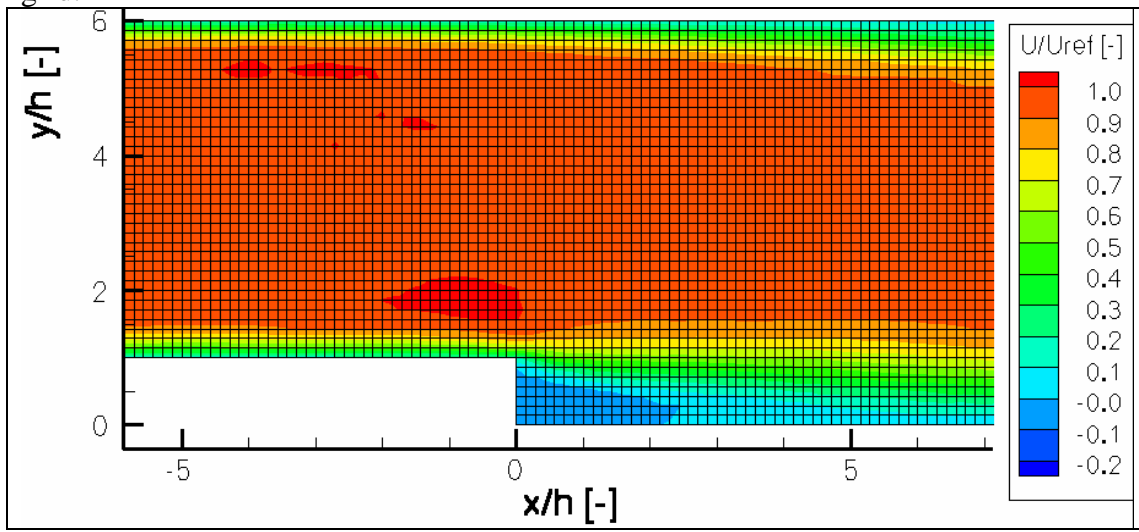

**Figure 1: Contour plot of normalized axial velocity and the macroscopic rigid ppr-grid (8330 ppr-cells)**

One can see that the ppr-grid is relatively coarse. The step is here resolved only with 7 ppr-cells and the whole domain has 8330 ppr-cells. It is important to keep in mind that the self-adaptive sub-grid (particle level on microscopic scale) on which the simulation is running is much finer. This validation cases used about 1 million particles. Regarding the computational times, the benchmarking of CFD2k is published in the user guide [19], which can also be downloaded on www.cfd2k.eu. In order to discuss the flow results in more detail Figure 2 shows instead of the ppr-grid the streamlines.

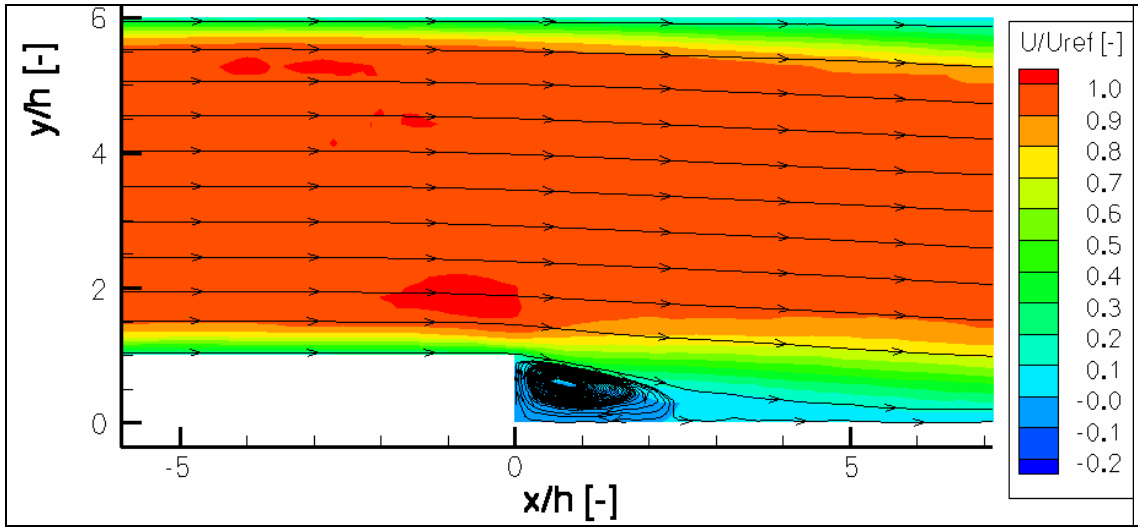

**Figure 2: Contour plot of normalized axial velocity with streamlines**

In Figure 2 one can see that of course a recirculation zone is build up behind the step and a shear flow between the wake and the bulk flow arises. The recirculation zone seems to be too small what is proven by a quantitative comparison with the ERCOFTAC dataset, which is shown in Figure 3.

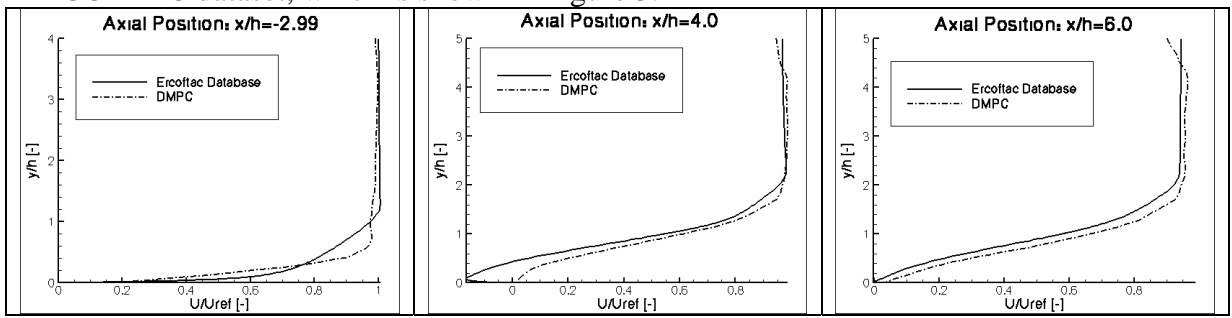

**Figure 3: Comparison of velocity profiles at 3 axial positions**

In the middle plot of Figure 3 where the velocity profiles at the axial position $x/h=4.0$ are shown, one can see that there are still negative velocities in the ERCOFTAC dataset what means that the recirculation zone is still existent at this axial position. In the DMPC result, no negative velocities occur any more, what indicates that the recirculation zone is over at this axial position. The reattachment location of the recirculation zone is given for the ERCOFTAC dataset to $(x/h)_{reattach}=6$. DMPC calculates for the reattachment coordinate of the recirculation zone $(x/h)_{reattach}=2.4$. After investigation about the reasons of this high discrepancy it was found out, that the modeling of the turbulent transport processes are afflicted with inaccuracies. The problem which DMPC had, is a to low particle density in wakes which is generated by wrong turbulent transport processes. One idea to solve this inaccuracy was to introduce corrections terms for the turbulent transport which work with particle insertion and particle merging depending on the local conditions. These functions of course ensure further on fully conservation of mass, momentum and energy and the results get better, but still not satisfying. This leads finally to the development of a particle model, which combines the advantages of an Euler-mesh method and a Lagrange-particle method. It is an approach using so-called moving Voronoi-cells. Most parts of the DMPC model stay identical, that is also the reason why the model name DMPC is kept also for the Voronoi approach. The difference of the model conception between the overlapping circular particles and the (non-overlapping) moving Voronoi-cells is visualized in Figure 4.

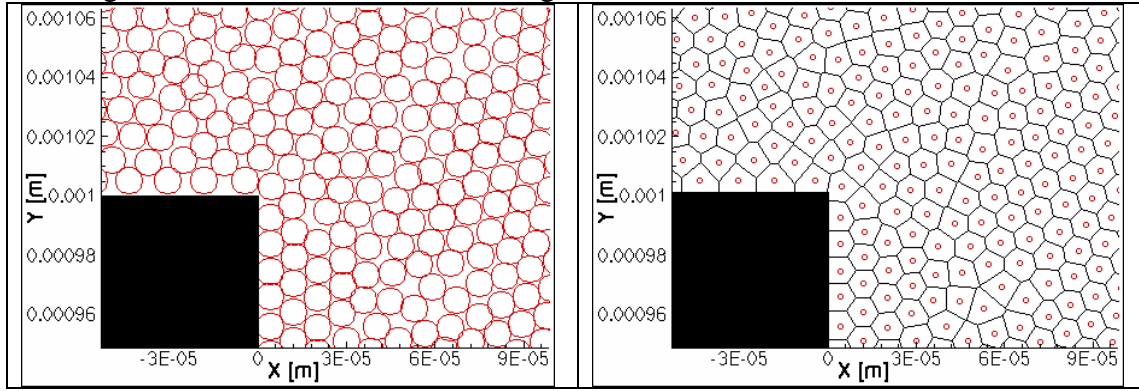

**Figure 4: Model conception of the sub-grid particles: Left: round overlapping particles. Right: Not-overlapping moving Voronoi-particles**

Figure 4 shows a zoom inside a ppr-cell resolving the self-adaptive sub-grid particles. Animations of the movement of the sub-grid particles for both approaches are shown at www.cfd2k.eu. These animations are very helpful to improve the imagination how the model really works in detail. Voronoi-cells are in literature [20, 21] better known as Voronoi-diagrams or Dirichlet tessellations. A Voronoi-diagram is a special kind of decomposition of a metric space determined by distances to a specified discrete set of objects in the space [22]. In this case the objects are the particle centers (red circles in the right picture of Figure 4). In the simplest and most common case the definition of the Voronoi diagram is as follows [23]: if in the plane a set of points N are given, than the Voronoi diagram for N is the partition of the plane which associates a region R(p) with each point p from N in such a way that all points in R(p) are closer to p than to any other point in N. In this approach the set of points N are the particle centers or Voronoi-cell centers (red circles in the right picture of Figure 4).

This approach uses moving Voronoi-cells i.e. at every time step the Voronoi-diagram has to be calculated again. But the information of previous time step can be used and there is a very efficient algorithm implemented which ensures more or less same computational needs for the Voronoi-approach compared to the circular particle approach. Of course the position, shape and area of every Voronoi cell is changing in every new time step. This changes are induced e.g. by a compression (pressure wave) but of course also by normal convection of the particles. An important advantage of the Voronoi approach is that the whole domain is consistently covered by these particles i.e. it is a kind of smoothed distribution compared to the circular particle approach. The moving-Voronoi approach can be regarded as an intermediate piece between an Euler grid approach and a mesh-less particle approach, which uses the advantage of both. The problems with the correct description of the turbulent transport processes can be solved much easier compared to the circular particle approach. This model extension is so far implemented in a non-released beta-version of the test-code CFD2k. The results are very promising and will be published after the upgraded version of CFD2k is fully validated and released.

## Summary

DMPC is a fully compressible flow model, which uses the Lagrange method i.e. moving particles to describe the flow field. Due to the multiple particle interactions there are exchanges of mass, momentum and energy between the involved particles and boundaries. In the first model version all particles have the same volume and a circular shape. Due to convection of the particles, an overlap is possible. This approach was afflicted with inaccurate turbulent transport, which leads e.g. to too small recirculation zones. An upgrade of the model to the moving Voronoi-cell approach ensures a much more accurate modelling of the turbulent transport processes, which finally lead to very promising results. For correct understanding of the model it is important to point out that the particles represent artificial flow elements, which posses certain flow variables and thermodynamic characteristics (dissipative and dispersive). The particles or the moving Voronoi-cells can also be interpreted as a self-adapting sub-grid (micro-scale). Adapting thermodynamic laws and similarity hypothesis as well as statistical laws over a sample of particles allows deriving averaged flow field data, which are stored in so-called macroscopic ppr-cells. A test-code called CFD2k is developed and published on www.cfd2k.eu. The released version of CFD2k is V0.1.12, which incorporates the circular overlapping particle approach. CFD2k can be downloaded on the web-page as well as the user guide and several test cases. The model upgrade using the (non-overlapping) moving Voronoi-cell approach is so far only implemented in an unreleased beta version of CFD2k (V0.1.13-beta). As soon as this new release will be fully validated, it will be set free for download and the detailed model description together with a validation report will be published.